\documentclass[preprint]{aastex}

\shorttitle{Time dependence of the proton flux}
\shortauthors{Adriani et al.}

\begin{document}

\title{Time dependence of the proton flux measured by PAMELA
   during the July 2006 - December 2009 solar minimum.} 

\author
{O. Adriani$^{1,2}$, G. C. Barbarino$^{3,4}$,
  G. A. Bazilevskaya$^{5}$, R. Bellotti$^{6,7}$, 
M. Boezio$^{8}$, \\
E. A. Bogomolov$^{9}$, M. Bongi$^{1,2}$, V. Bonvicini$^{8}$,
S. Borisov$^{10,11,12}$, S. Bottai$^{2}$, \\
A. Bruno$^{6,7}$, F. Cafagna$^{7}$, D. Campana$^{4}$,
R. Carbone$^{8}$, P. Carlson$^{13}$, \\
M. Casolino$^{11}$, G. Castellini$^{14}$
M. P. De Pascale$^{10,11,\dagger}$, C. De Santis$^{10,11}$, N. De
Simone$^{10}$,\\  
V. Di Felice$^{10}$, V. Formato$^{8,15}$, 
 A. M. Galper$^{12}$, L. Grishantseva$^{12}$,
A. V. Karelin$^{12}$, \\
S. V. Koldashov$^{12}$, S. Koldobskiy$^{12}$, 
S. Y. Krutkov$^{9}$, A. N. Kvashnin$^{5}$,
 A. Leonov$^{12}$, \\
V. Malakhov$^{12}$, 
L. Marcelli$^{11}$, 
A. G. Mayorov$^{12}$, W. Menn$^{16}$, V. V. Mikhailov$^{12}$, \\
E. Mocchiutti$^{8}$, 
A. Monaco$^{6,7}$,  N. Mori$^{2}$, N. Nikonov$^{9,10,11}$, G. Osteria$^{4}$,
F. Palma$^{10,11}$, \\
P. Papini$^{2}$, M. Pearce$^{13}$, P. Picozza$^{10,11}$, 
C. Pizzolotto$^{8,17,18}$ M. Ricci$^{19}$,
S. B. Ricciarini$^{14}$, \\
L. Rossetto$^{13}$, R. Sarkar$^{8}$, M. Simon$^{16}$,
 R. Sparvoli$^{10,11}$, P. Spillantini$^{1,2}$, \\
Y. I. Stozhkov$^{5}$, A. Vacchi$^{8}$, 
E. Vannuccini$^{2}$, G. Vasilyev$^{9}$, S. A. Voronov$^{12}$, \\
 Y. T. Yurkin$^{12}$, J. Wu$^{13,*}$,
 G. Zampa$^{8}$, N. Zampa$^{8}$, 
 V. G. Zverev$^{12}$, \\
M. S. Potgieter$^{20}$, E. E. Vos$^{20}$}
\affil{$^{1}$University of Florence, Department of Physics, I-50019 Sesto Fiorentino, Florence, Italy}
\affil{$^{2}$INFN, Sezione di Florence, I-50019 Sesto Fiorentino, Florence, Italy}
\affil{$^{3}$University of Naples ``Federico II'', Department of Physics, I-80126 Naples, Italy}
\affil{$^{4}$INFN, Sezione di Naples,  I-80126 Naples, Italy}
\affil{$^{5}$Lebedev Physical Institute, RU-119991, Moscow, Russia}
\affil{$^{6}$University of Bari, Department of Physics, I-70126 Bari, Italy}
\affil{$^{7}$INFN, Sezione di Bari, I-70126 Bari, Italy}
\affil{$^{8}$INFN, Sezione di Trieste, I-34149 Trieste, Italy}
\affil{$^{9}$Ioffe Physical Technical Institute,  RU-194021 St. Petersburg, Russia}
\affil{$^{10}$INFN, Sezione di Rome ``Tor Vergata'', I-00133 Rome, Italy}
\affil{$^{11}$University of Rome ``Tor Vergata'', Department of Physics,  I-00133 Rome, Italy}
\affil{$^{12}$Moscow Engineering and Physics Institute,  RU-11540 Moscow, Russia}
\affil{$^{13}$KTH, Department of Physics, and the Oskar Klein Centre for Cosmoparticle Physics, AlbaNova University Centre, SE-10691 Stockholm, Sweden}
\affil{$^{14}$IFAC, I-50019 Sesto Fiorentino, Florence, Italy}
\affil{$^{15}$University of Trieste, Department of Physics, I-34147 Trieste, Italy}
\affil{$^{16}$Universit\"{a}t Siegen, Department of Physics, D-57068 Siegen, Germany}
\affil{$^{17}$INFN, Sezione di Perugia, I-06123 Perugia, Italy}
\affil{$^{18}$Agenzia Spaziale Italiana (ASI) Science Data Center, I-00044 Frascati, Italy}
\affil{$^{19}$INFN, Laboratori Nazionali di Frascati, Via Enrico Fermi 40, I-00044 Frascati, Italy}
\affil{$^{20}$Centre for Space Research, North-West University, 2520 Potchefstroom, South Africa}
\affil{$^{*}$Now at School of Mathematics and Physics, China University of Geosciences, CN-430074 Wuhan, China}
\affil{$^{\dagger}$Deceased}

\begin{abstract}
The energy spectra of galactic cosmic rays carry fundamental information
regarding their origin and propagation. These spectra, when measured near
Earth, are significantly affected by the solar magnetic field. A
comprehensive description of the cosmic radiation must therefore include
the transport and modulation of cosmic rays inside the heliosphere. During
the end of the last decade the Sun underwent a peculiarly long quiet phase
well suited to study modulation processes. In this paper we present proton
spectra measured from July 2006 to December 2009 by PAMELA. The large
collected statistics of protons allowed the time variation to be followed on
a nearly monthly basis down to 400 MV. Data are compared with a
state-of-the-art 
three-dimensional model of solar modulation.
\end{abstract}


\section{Introduction}

Protons are the most abundant species in cosmic rays, representing
about 90\% of the total flux.  
They are believed to be mainly of primordial origin, accelerated during 
supernovae explosions and confined within our Galaxy, at least up to
PeV energies.  
During propagation through the interstellar medium cosmic rays undergo
interactions with gas atoms  and loose energy
significantly modifying their injection spectra and composition. 

Beside the effect of propagation in the Galaxy, cosmic-ray
particles reaching the Earth are affected by the solar wind and the
solar magnetic field. 
A continuous flow of plasma emanates from the solar corona and extends
out far beyond Pluto, up to the heliopause where the interstellar
medium is encountered. The solar wind travels at supersonic speeds of
about 350-650 km~s$^{-1}$ up to the termination shock where it
slows down to about 100-150 km~s$^{-1}$ to form the inner heliosheath. 

The plasma carries the solar magnetic field out into the heliosphere
to form the heliospheric magnetic field (HMF). When interstellar
cosmic rays enter the heliosphere, they interact with the solar wind
and the HMF that modify their energy spectra. These
modifications are directly related to  solar activity and the total
effect is called solar modulation. Solar activity varies strongly with
time, rising from a minimum level when the Sun is quiet (with cosmic
rays then having a maximum intensity at Earth) to a maximum period
(when cosmic rays have a minimum intensity) and then returning to a
new minimum to repeat the cycle. This solar cycle spans
approximately 11 years. Solar activity modulates cosmic rays with the
same cycle as a function of position in the heliosphere.
Protons with
rigidities up to at least 30 GV are affected and the effect becomes
progressively 
larger as the rigidity decreases (e.g. \citet{pot11}).  
At each solar activity maximum
the polarity of the solar magnetic field reverses. When the solar
magnetic field lines point outward in 
the Northern hemisphere and inward in the Southern hemisphere, the Sun
is said to be in an A$>$0 cycle, whereas during an A$<$0 cycle the
solar magnetic field lines point inward (outward) 
in the Northern (Southern) hemisphere.
Direct cosmic-ray measurements above a few MeV have been performed
from the inner to the outer heliosphere (see e.g. \citet{heb06,web12})
and particularly at the Earth where solar modulation effects are at
their largest.  

Understanding the 
origin and propagation of cosmic rays requires knowledge of the
cosmic-ray energy 
spectra in the interstellar  
medium, i.e. uninfluenced by the Sun's magnetic field. These spectra,
especially for   
protons, can be inferred from gamma-ray data
(e.g. see~\citet{ner12}). However, uncertainties are large and  
direct cosmic-ray measurements are preferable. Voyager space missions
\citep{dec05,ric08} have passed the  
termination shock and may soon start sampling the cosmic radiation in
the local interstellar  
medium. They will provide fundamental data to understand the basic
features of the heliosphere. However, these cosmic-ray  
measurements are limited to the very low energy part of the spectra
(from tens to a few 100 MeV), with galactic protons, helium and oxygen
strongly contaminated by the anomalous component \citep{sto08}. Higher
energies  
are needed to fully address fundamental questions in cosmic-ray physics. 
Moreover, the energy spectra of rare components like antiparticles may
yield indirect information regarding new 
physics such as dark matter. For example, the antiproton flux between
few hundred  
MeV- few GeV can be used (e.g. \citet{cer12,asa12,gar12}) to constrain
models with light ($\sim 10$~GeV) dark matter  
candidates put forward to interpret recent measurements by
direct-detection experiments \citep{bern00,aal11,ang12}. 
However, solar modulation has a relevant effect on the cosmic-ray
spectra at these energies that needs to be  
disentangled to allow a comprehensive picture to emerge.

Precise measurements of cosmic-ray spectra over a wide rigidity range,
from a few hundred MV to tens of GV over an extended period of time
(covering a significant fraction of the solar cycle) can be used to
study 
the effect of solar modulation in greater detail, including 
the convective and adiabatic cooling effect of the expanding solar
wind and the diffusive and particle drift effects of the turbulent
HMF. 
The most recent solar minimum activity period (2006-2009) and the
consequent 
minimum modulation conditions for cosmic rays were unusual. It was
expected that the new activity cycle would begin early in
2008. Instead minimum modulation conditions continued until the end of
2009 when the highest levels of cosmic rays were recorded at Earth
since the 
beginning of the space age (see e.g. \citet{mew10}). This period
of prolonged solar minimum activity is well suited to study the
various modulation processes that affect the propagation of galactic
cosmic rays inside the heliosphere. 

PAMELA is a satellite-borne instrument designed for cosmic-ray
antimatter studies. 
The instrument is flying on-board the Russian Resurs-DK1 satellite
since June 2006, in a semi-polar  
near-Earth orbit. PAMELA has provided important results on the
antiproton~\citep{adr09a} and  
positron~\citep{adr09b} galactic abundances.
The high-resolution magnetic spectrometer allowed hydrogen and helium
spectral measurements up  
to 1.2 TV~\citep{adr11}, the highest limit ever achieved by this kind
of experiment. We present here  
results on the long-term variations in the absolute flux of hydrogen
nuclei, measured down to 400~MV.  


The analysis, described in section~\ref{sec:analysis}, is based on
data collected between July 2006 and December 2009. 
The analyzed period covers the unusually long most recent solar
minimum.  
The large proton
statistics collected by the instrument allowed 
the proton flux to be measured for each Carrington
rotation\footnote{Mean synodic rotational period of the Sun surface,
  corresponding to 
about 27.28 days~\citep{car63}.} (hereafter the flux measured on this
time basis will be referred as Carrington flux).

\section{The instrument}

\begin{figure}[h]
\begin{center}
\includegraphics[width=0.5\textwidth]{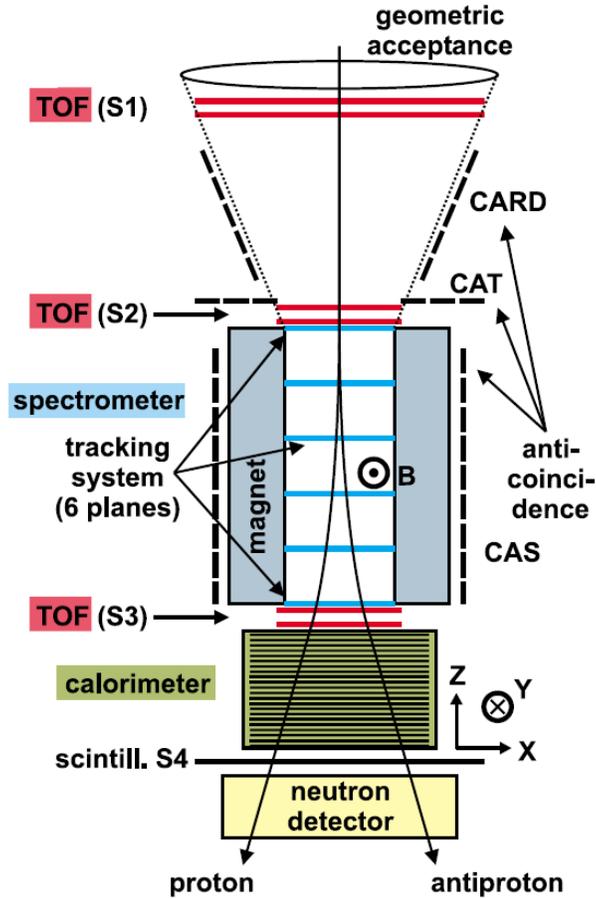}
\end{center}
\caption{Schematic view of the PAMELA apparatus.\label{fig:appa}}
\end{figure}

The PAMELA instrument, shown schematically in Figure~\ref{fig:appa},
comprises a number of highly redundant  
detectors, capable of identifying particles by providing charge
($Ze$), mass, rigidity  
($R = pc/Ze$, with $p$ momentum of the particle and $c$ speed of
light) and velocity ($\beta$) measurements over  
a very wide energy range. Multiple detectors are placed around a
permanent magnet with a silicon micro-strip tracking  
system, providing charge and track deflection ($ d = \pm 1/R$)
information. A scintillator system (TOF: S1, S2 and S3)  
provides the trigger for event acquisition, the time-of-flight
measurement and a further charge information.  
A silicon-tungsten calorimeter is used to perform hadron/lepton
separation; a scintillator (S4) and a neutron detector  
placed at the bottom of the apparatus increase the rejection power at
high energy. An anti-coincidence system (AC: CAS, CAT and CARD)  
registers hits produced by particles entering from outside the
instrument acceptance. A detailed description of the  
apparatus can be found in \citet{pic07}. 

\section{Analysis}
\label{sec:analysis}

\subsection{Event selection}

In order to give a valid trigger for data acquisition, a particle must
cross the S2 and S3 scintillators, located above  
and below the tracking system, respectively. The requirement of a hit
on S3 sets a 
limit on the minimum   
detectable rigidity, which for vertical protons is on average 400~MV,
i.e. $\simeq 80$~MeV of kinetic energy. 

Clean events are selected by requiring a precision 
track in the spectrometer, a trajectory contained within the 
instrument acceptance and no hits in the CARD and CAT
scintillators. Protons are identified  
among all the selected particles by requiring a positive track
curvature and
ionization energy losses (dE/dX) in the 
tracking system consistent with a hydrogen nucleus. The dE/dX
selection included deuterons, which in this work  
have been treated as being protons. Galactic particles have been
selected by requiring that the measured  
rigidity was a factor $1.3$ above the geomagnetic vertical cutoff,
evaluated using the St\"{o}rmer approximation, at a given orbital
position. The  
satellite orbit has an inclination of 70$^{\circ}$, which allows
galactic particles to be studied down to  
the minimum detectable rigidity. 

The analysis procedure was similar to previous work on the high energy
proton and helium fluxes~\citep{adr11}.  
The major differences concern track-quality requirements and the
fiducial volume definition. The procedure used in the proton and
helium study aimed to maximise 
the spectrometer performance and minimize systematic uncertainties in
the flux normalization,  
in order to perform a precise spectral measurement up to the highest
achievable energy. The study presented in this paper aims to trace
time variations in the low energy particle flux, with the finest
possible time binning. Hence, high-performing tracking was 
less important than high statistics. For this reason the tracking
requirements were relaxed (at least 3 
hits, instead of 4, on both X and Y view, a track lever-arm of at
least 4 silicon planes in the tracker)   
and the fiducial volume was enlarged (bounded 1.5~mm from the
magnet cavity walls, instead of 7~mm), allowing to collect  
more particles with only a small degradation in the measurement
precision in the energy range under study.  

\subsection{Flux normalization}

The absolute proton flux in the spectrometer was obtained by dividing
the measured energy spectrum by the acquisition time,  
the geometrical acceptance and the selection efficiency.

\subsubsection{Acquisition time}

Flux evaluation was performed as a function of time, binning the data
according to the Carrington rotations. The 
time period studied 
corresponds to Carrington rotation numbers 2045-2092. The acquisition
time was evaluated for each energy bin  
as the total live time spent above the geomagnetic cutoff, outside the
South Atlantic Anomaly and during periods when no solar  
events had occurred (hence December 2006, when a large solar event
took place, was excluded from the data). This work is  
based on data collected between July~2006 and December~2009. The total
live time was about 5$\times 10^{7}$s ($\sim$ 590~days) 
above $\sim$ 20~GV, reducing to about 13$\%$ of this value at 0.4~GV,
due to the relatively short time spent by the satellite  
at low geomagnetic locations. The accuracy of the live time
determination was cross-checked by comparing different clocks  
available in flight, which showed a relative difference of less than
0.2~$\%$. 

\subsubsection{Geometrical factor}

The geometrical factor for this analysis ($G=$19.9~cm$^2$sr) is
determined by the requirement of containment within the fiducial  
volume: a reconstructed track had to be 1.5~mm away from the magnet
walls and the TOF-scintillator edges. The geometrical factor is
constant  
above 1~GV and slowly decreases by $\sim$ 2\% to lower energies, where
the curvature of the proton particle trajectory is no longer 
negligible. 

The value of $G$ has been evaluated with two independent numerical
methods, which provided values consistent within 0.1$\%$. 

\subsubsection{Selection efficiency}
The PAMELA instrument has been designed to provide redundant
information on event topology and particle characteristics. This 
feature allowed the efficiencies of each detector to be measured
directly from flight data. Due to the large proton statistics,
detector behavior could be monitored over very short time scales. For
this analysis, efficiencies were 
evaluated independently for each time interval, in order to follow any
possible time variation of detector responses. 

The major time dependent effect was the sudden failure of some
front-end chips in the tracking system. 
Chip failures started to occur few months after the launch of the
satellite, initially 
with an almost constant rate. In the following years the
rate of failures significantly decreased and the tracking system
reached an almost stable  
condition near the end of 2009. 
The net effect was a progressive reduction of
the tracking efficiency, since the number of hits available for track
reconstruction decreased. No  degradation in the signal-to-noise ratio
and spatial resolution was observed.

The approach described in \citet{adr11} for the evaluation of the
tracking efficiency  
was used: a high energy average value was experimentally obtained
using 
non-interacting minimum ionizing particles in 
the calorimeter from flight data   
and an energy dependent correction factor  
was evaluated using Monte Carlo simulation. The average tracking
efficiency was measured using a test sample of  
non-interacting relativistic protons in the imaging calorimeter. The
resulting efficiency varied almost monotonically from  
$\simeq$ 94$\%$ during the first months to $\simeq$ 20$\%$ in the final 
months. 
The efficiency variation was correctly reproduced by the simulation,
which included a map of dead channels as a  
function of time. Since the requirement of non-interacting minimum
ionizing particles in the calorimeter selected 
essentially protons with rigidities $ \geq 2$ GV, the determination of
the tracking efficiency at lower rigidities was performed  
using the simulation.
The simulation results were normalized to experimental values at high
rigidities after performing an identical selection.  
This was required because, due to 
the relatively poor tracking resolution of the calorimeter, the average
efficiency values obtained using  
the calorimeter selection were different from the average efficiency
of an isotropically-distributed particle sample (as was 
the case for triggered events). This normalization was applied to the
simulation estimated efficiency that accounted for the  
full energy dependence of the tracking efficiency and 
was used for the flux estimation. When comparing the average 
experimental and simulated efficiencies, differences of less than 4\%
were found, except during the last few months 2009 when the tracking  
efficiency fell below 40\% and, consequently, the discrepancy became
more significant, increasing up to 13\%. These differences  
were included in the systematic uncertainties of the measurements. 
As a further check, two different simulation packages were used:
GEANT3~\citep{bru94} and GEANT4~\citep{ago03}. The resulting
efficiencies differed by $\simeq 1.5\%$ and this was included in the
systematic uncertainty.  

TOF and AC selection efficiencies were evaluated relative to the
tracking selection. For well contained events, the TOF and AC  
responses were not correlated and the selection efficiencies could  be
measured independently. At low rigidities
TOF and AC efficiency measurements were affected by   
background contamination in the test samples of low energy secondary
particles produced by high energy primaries interacting with the
apparatus.  
This contamination resulted in an underestimation of the efficiency,
since background interactions were often associated with multiple hits
in both the TOF and AC scintillators. 
An upper limit on this contamination was determined by excluding the
TOF and AC conditions from the test sample selection, thereby 
adding background events into the test sample. The difference between
the results was found to be $\simeq$ 6$\%$ in the lowest-energy bin. This
value  
was conservatively treated as an asymmetric systematic uncertainty on 
the absolute flux. 

\subsection{Spectral unfolding and other corrections}

The proton spectrum measured in the spectrometer was corrected to the
top of the payload by accounting for the effects of rigidity  
displacement due to finite spectrometer resolution and particle
\begin{figure}[h]
\begin{center}
\includegraphics[width=0.8\textwidth,height=0.2\textheight]{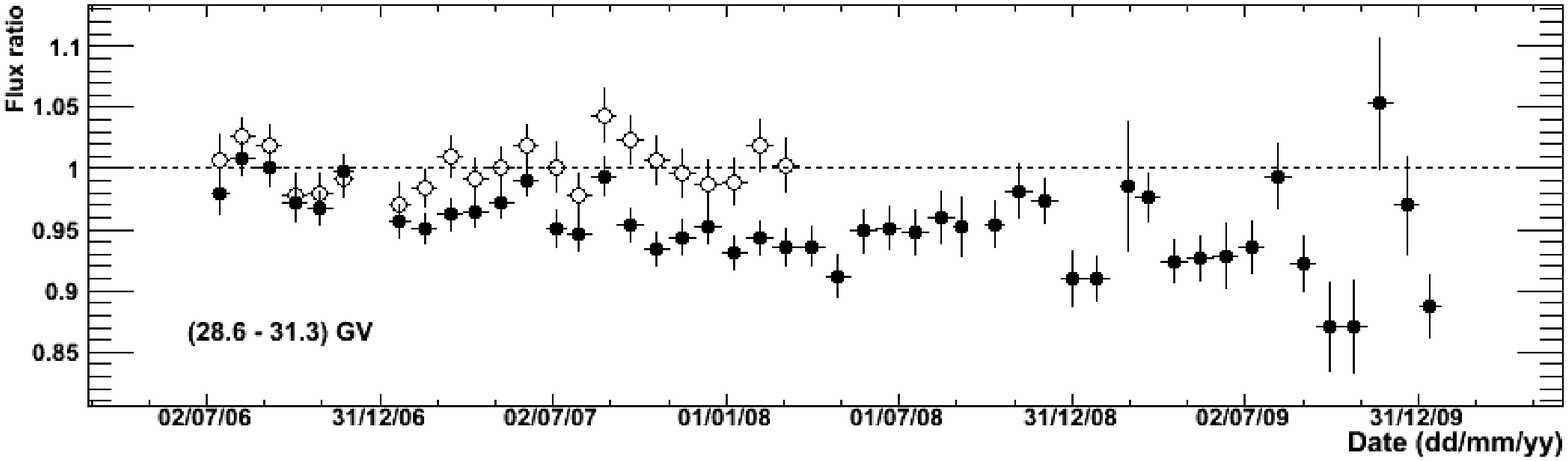} 
\includegraphics[width=0.8\textwidth,height=0.2\textheight]{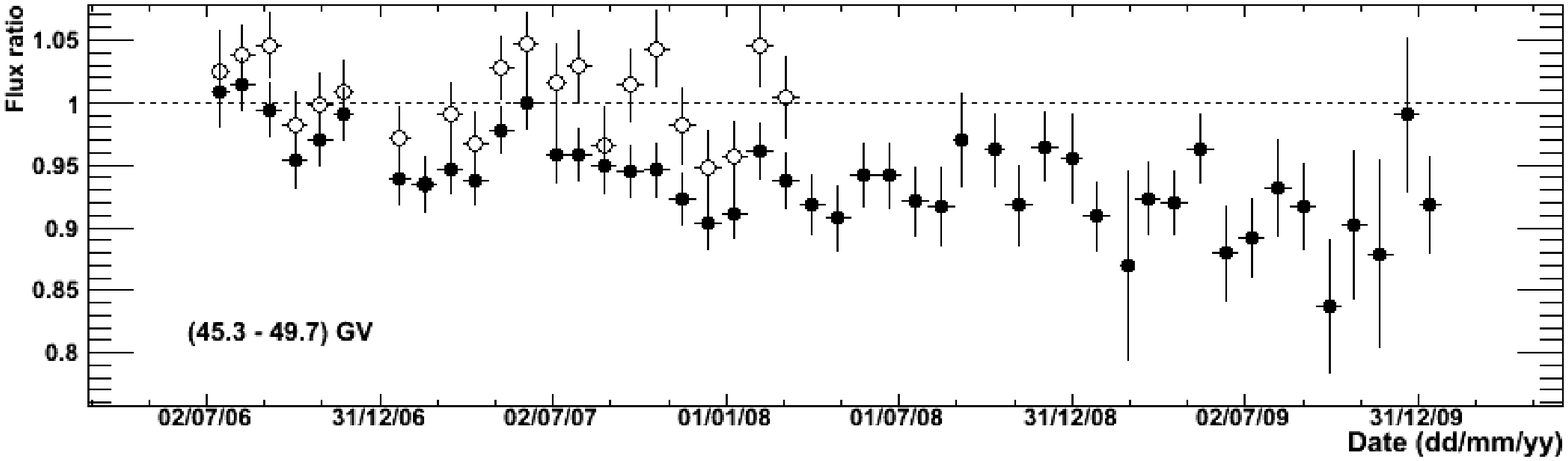} 
\end{center}
\caption{Carrington proton fluxes shown relative to the flux obtained as
  in \citet{adr11} (value averaged
  over the period July 2006 - March 2008)  
  for (open circles) stringent tracking criteria and (filled
  circles) relaxed tracking criteria. Two energy intervals are presented.
\label{fig:timeprof} }
\end{figure}
slow-down by ionization energy loss, the latter being more  
relevant at low energies. The correction was applied by means of an
unfolding procedure, following  
the Bayesian approach proposed by \citet{dag95}. The detector response
matrix was obtained from the simulation. The resulting  
spectrum was corrected for the contamination of locally produced
secondary particles and for attenuation due to particles  
interacting with the satellite and the instrument itself. 
In order to estimate the residual contamination of locally produced
secondary particles, simulations of protons and helium 
nuclei isotropically hitting the PAMELA payload within a 2$\pi$ solid
angle were 
carried out. The estimated background was found to be less than 6$\%$ at   
0.4~GV and inversely proportional to energy, 
becoming negligible above 1~GV. Flux attenuation, which was slightly
energy  
dependent and amounted to 5-6$\%$, accounted for both particle loss by
non-elastic interactions and loss and gain of particles from within
the acceptance due to elastic scattering.

Finally, in order to correct for systematic time-dependent variations
of the 
Carrington fluxes (see next
section), these  
were normalized, at high-energy (30-50 GeV), to the proton flux measured 
with the more stringent 
tracking conditions of \citet{adr11} 
and averaged over the period
July 2006 - March 
2008\footnote{An improved use of the
  simulation resulted in slightly higher selection efficiencies than
  previously estimated.
Therefore,  
the proton spectrum of \citet{adr11}
should be lowered by 3.2\%. Furthermore, 
the 
overall systematic uncertainty 
is slightly larger at low energies (4.5$\%$ at 1~GV).}.

\subsection{Systematic uncertainties}

By relaxing the tracking selection the proton statistics could be
significantly increased, allowing the proton flux to be evaluated on  
a Carrington rotation 
basis up to the end of 2009. This approach also led to an
increase of the systematic uncertainties on the measured  
absolute fluxes. 

The reliability of the flux determination was verified by studying the
time variation of the flux at high energy (above 30 GeV) 
since no significant modulation effects are expected in this energy
region. As a reference, we considered the fluxes obtained with the
stringent tracking conditions and 
averaged over the period 
July 2006 - March 2008. Figure~\ref{fig:timeprof} shows the time
dependence of the 
ratio between 
the Carrington fluxes and the reference flux for two high-energy
intervals at about 30 GV 
and 47 GV. Two different selection criteria were considered:
stringent 
(open circles, as in \citet{adr11})
and relaxed tracking selections 
(filled circles, this work). 

Comparing the upper and 
lower panels of Figure~\ref{fig:timeprof} a correlation between the
time profiles at different energies can be noticed, which  
indicates the presence of non-negligible systematic effects. 
The Carrington fluxes based on the 
analysis described in \citet{adr11} 
\begin{figure}[ht]
\begin{center}
\includegraphics[width=0.8\textwidth,height=0.2\textheight]{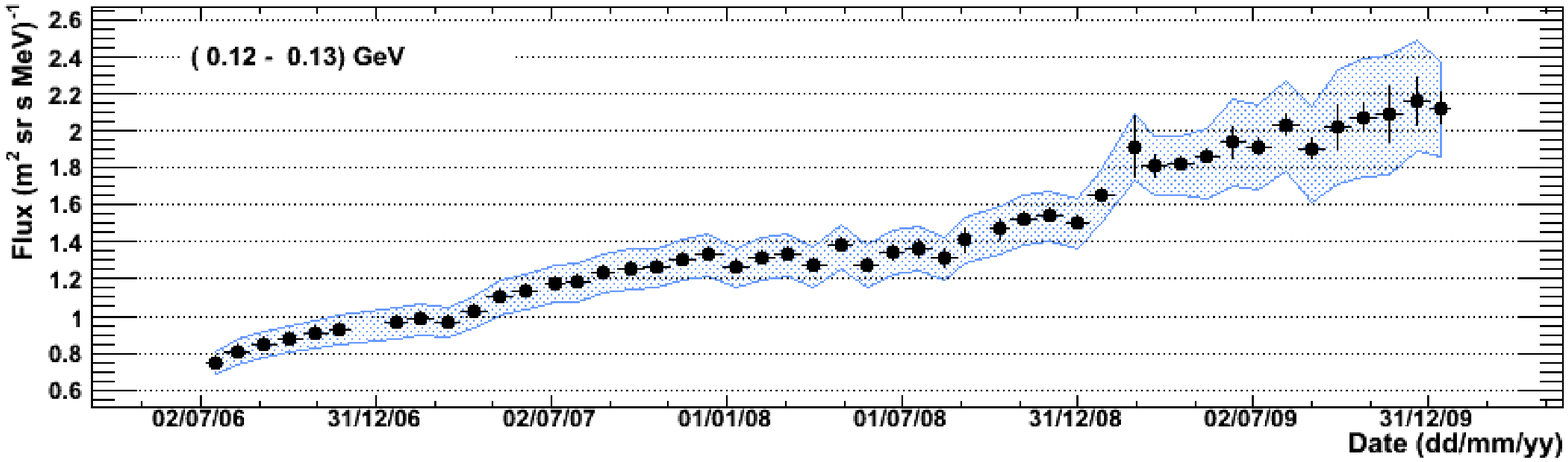} 
\includegraphics[width=0.8\textwidth,height=0.2\textheight]{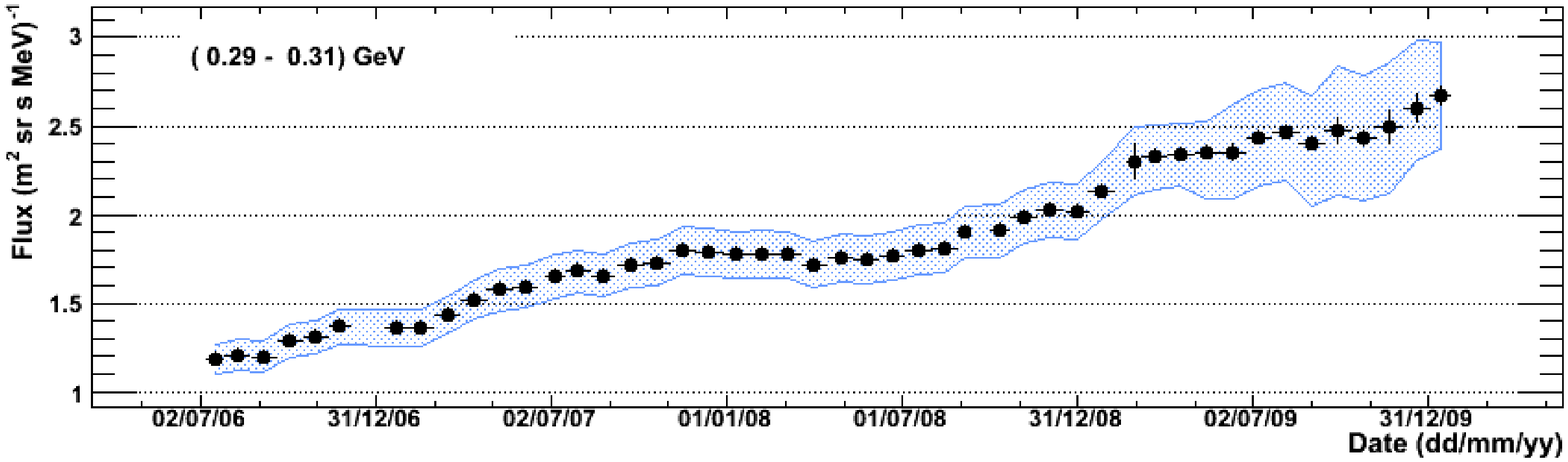} 
\includegraphics[width=0.8\textwidth,height=0.2\textheight]{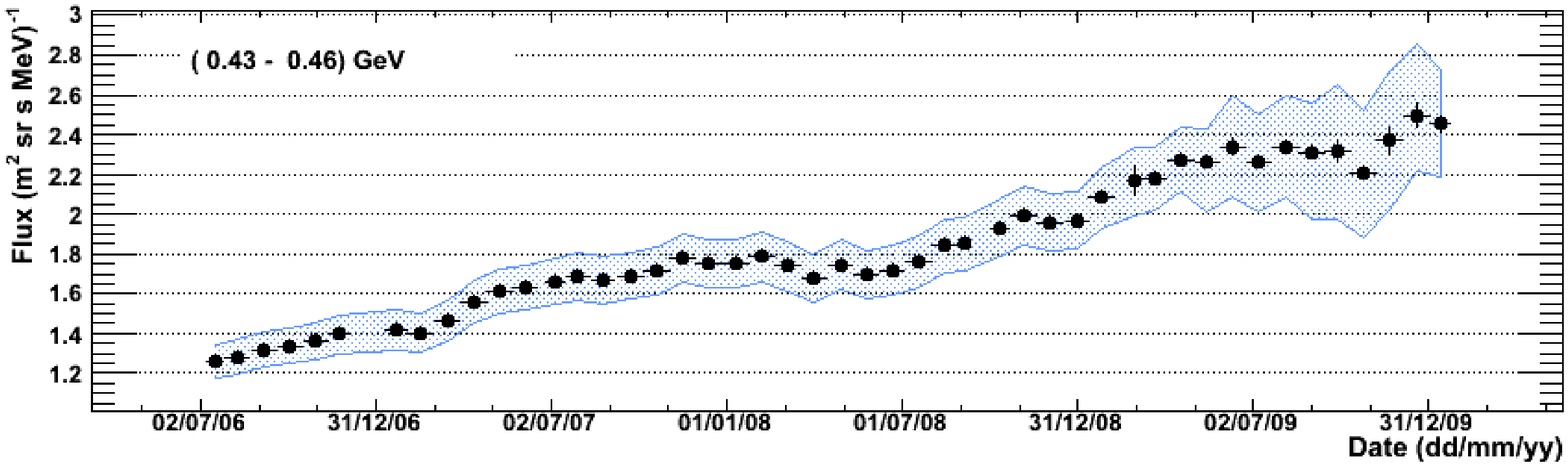} 
\end{center}
\caption{The time profile of the proton flux for three energy intervals (from
  top to bottom): 0.597-0.625~GeV, 0.99-1.04 GeV and
  2.99-3.13~GeV. The error bars are statistical and the shaded area
  represents the quadratic sum of all systematic uncertainties. 
\label{fig:timeprofnorm}}
\end{figure}
fluctuate around a mean value consistent with the 
quoted $\simeq 4\%$ systematic uncertainty, while the fluxes
for this  
analysis show a decreasing trend, with a relative variation over the  
whole time interval of $\sim $10\%. This time-dependent systematic
effect was corrected as discussed in the previous section.
The assumption behind this correction implies
that PAMELA results are not sensitive to solar modulation induced flux
variations smaller than about 5\%.

Further contributions to the total uncertainty resulted from the
unfolding procedure (1-5\%) and from the correction to  
the top of the payload (1$\%$).

\section{Results}

Figure~\ref{fig:timeprofnorm} shows the
time profile of the proton flux for three energy intervals; the error bars
are statistical, while the shaded area represents the quadratic sum of
all systematic uncertainties.  
The systematic uncertainty is dominated by the tracking efficiency and
reaches a minimum above a few tens of GeV and for the first Carrington
rotations, when the tracking efficiency is maximum.  
Additional time and energy-dependent contributions to the overall
uncertainty have been introduced in order to account for both the
reduced accuracy of the simulation in reproducing the spectrometer
behavior as the detection efficiency decreases in time, and the larger
uncertainty on the simulated efficiency at low energy. 

\begin{figure}[ht]
\begin{center}
\includegraphics[width=0.6\textwidth]{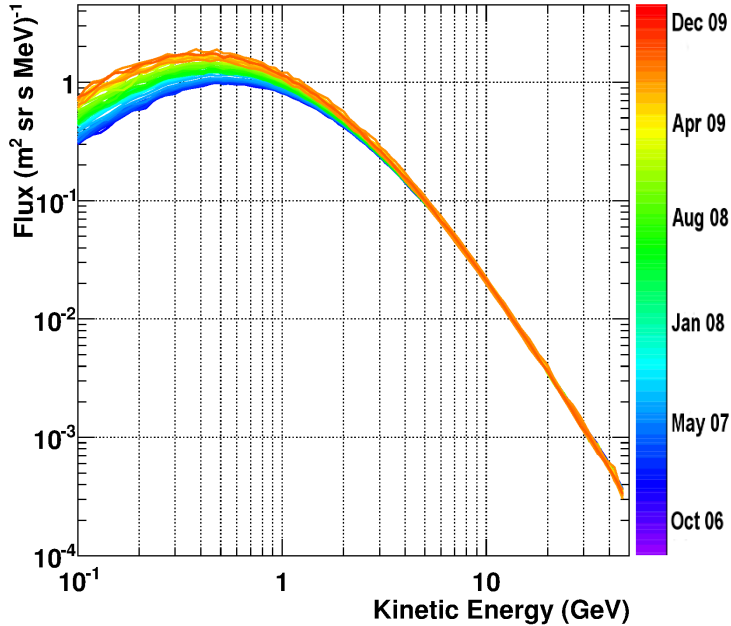} 
\end{center}
\caption{The evolution of the proton energy spectrum 
as particle intensities
approached the period of minimum solar activity, from July 2006
(violet), to December  
2009 (red). The region between the blue and red curves indicates the
spread in proton fluxes 
during this time.
\label{fig:fluxabs}}
\end{figure}

Figure~\ref{fig:fluxabs} shows the evolution of the  
Carrington proton energy spectra from July 2006 (blue curve) to
December 2009 (red curve).  
Around 100 MeV proton intensities increased by $\simeq 20\%$ over 
twelve months, from December 2006 to December 2007, with a slightly
larger increase 
of  $\simeq 25\%$ from December 2007 to December 2008. For 2009, when 
solar minimum conditions continued to prevail throughout the
heliosphere, intensities increased by almost $40\%$. From these data
it is clear that 
protons (evident across the greater part of the energy spectrum)
reached maximum intensities at 
the end of 2009.

\section{Data Interpretation and Discussion}

\begin{figure}[ht]
\begin{center}
\includegraphics[width=0.6\textwidth]{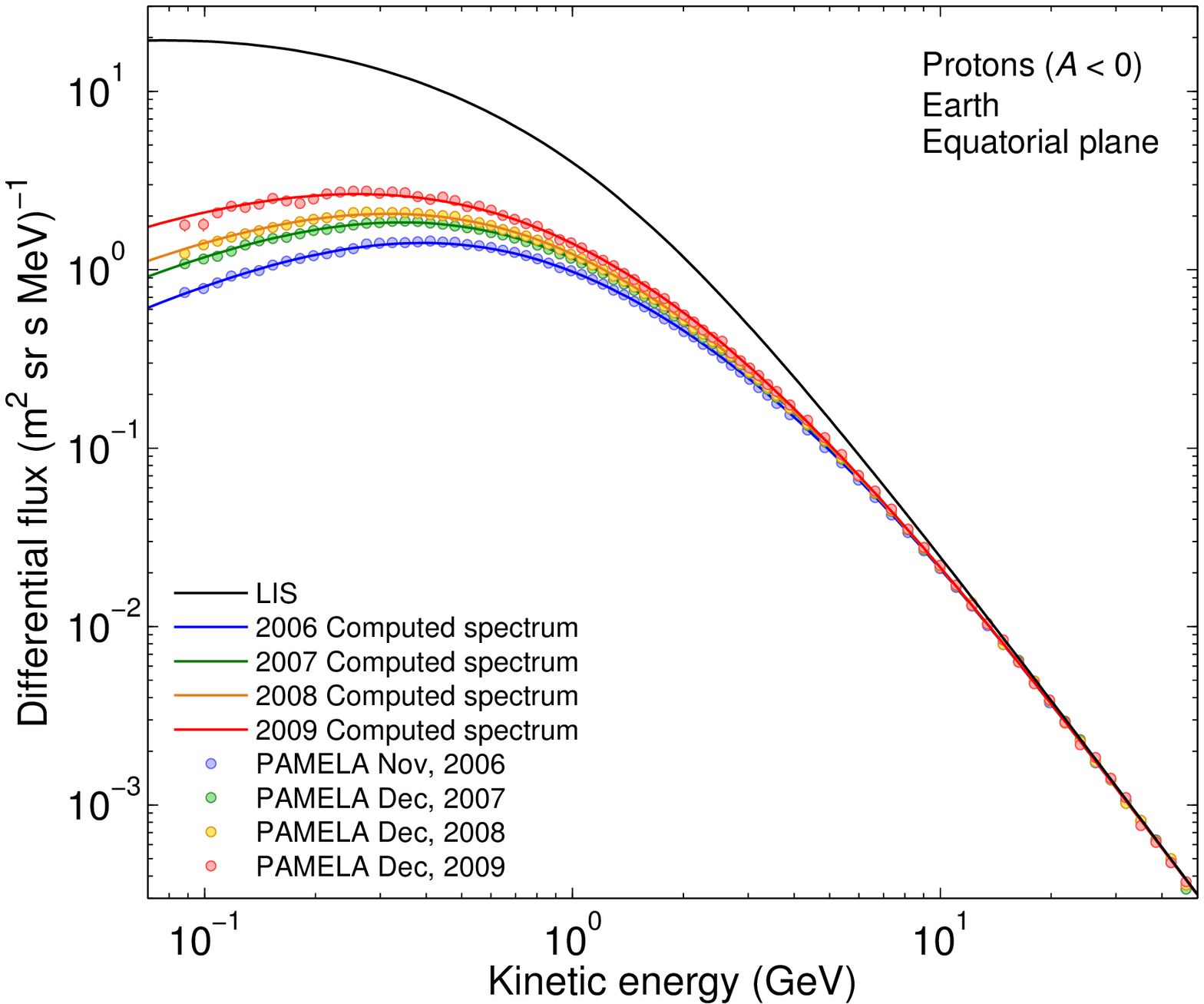} 
\end{center}
\caption{The PAMELA proton spectra for November 2006 (blue), December
  2007 (green), December 2008 (yellow) and December 2009 (red)
  overlaid with the corresponding computed spectra (solid lines) 
\citep{pot12}. 
  The LIS used for the computation is also shown (black solid line).
\label{fig:prot06071001andtheory}}  
\end{figure}

Figure~\ref{fig:prot06071001andtheory} and Table~\ref{tbl-1} 
present the galactic proton spectra
measured 
by PAMELA over four time periods. These data
illustrate how the proton spectra evolved from late 2006 to late
2009. The latter represents the highest proton flux observed by PAMELA. 
In Figure~\ref{fig:prot06071001andtheory}  
the PAMELA
proton spectra are overlaid with the corresponding computed 
spectra (solid lines).
A full three-dimensional model was used to compute the differential
intensity of cosmic-ray protons at Earth from 10 MeV to 30 GeV. The
model is based on the numerical solution of the Parker transport
equation~\citep{par65}, including all four major modulation
mechanisms: convection, particle drift caused by gradients and
curvatures in the HMF, diffusion described by a full 3D diffusion
tensor and adiabatic energy changes. For these calculations a local
interstellar 
spectrum (LIS) was assumed (shown as a solid black line in
Figure~\ref{fig:prot06071001andtheory}). The proton LIS was based
on that by \citet{lan04} modified at high
energies to match PAMELA observations. 
It follows from the comparison between the model and observations that
the proton spectra became progressively softer, as expected, from 2006
but requiring larger diffusion coefficients at lower
energies than anticipated. At the same time full particle drifts
occurred but with the modulation drift effects somewhat obscured by
the effective diffusion. This caused the effect of the HMF wavy
current sheet to be less well correlated to increases in cosmic-ray
intensities 
at Earth. Details concerning the modulation model and theoretical
assumptions and implications, will be published in an accompanying
paper \citep{pot12}. 

\acknowledgments

We acknowledge support from The Italian Space Agency 
(ASI), Deutsches Zentrum f\"{u}r Luft- und Raumfahrt (DLR), The
Swedish National Space  
Board, The Swedish Research Council, The Russian Space Agency
(Roscosmos) and The
Russian Foundation for Basic Research.
Partial financial support was given by the South African National
Research Council (NRF) for the numerical modeling .

\clearpage

\begin{deluxetable}{lllll}
\tabletypesize{\scriptsize}
\rotate
\tablecaption{Hydrogen flux measured by PAMELA over five time
  periods. The first and second errors represent the statistical and
  systematic uncertainties, respectively. \label{tbl-1}}
\tablewidth{0pt}
\tablehead{
\multicolumn{1}{c}{Kinetic Energy } & \multicolumn{4}{c}{Flux } \\ 
\multicolumn{1}{c}{(GeV)  } & \multicolumn{4}{c}{(m$^2$ s sr MeV)$^{-1}$}\\ \cline{2-5} 

                           &  \multicolumn{1}{l}{2006/11/13 - 2006/12/4} & \multicolumn{1}{l}{2007/11/30-2007/12/27} 
                           & \multicolumn{1}{l}{2008/11/19-2008/12/15} & \multicolumn{1}{l}{2009/12/6-2010/1/1}

}
\startdata

0.082 - 0.095& (0.721 $\pm$ 0.032 $^{+ 0.061}_{-0.100}$)		   & (1.045 $\pm$ 0.031 $^{+ 0.090}_{-0.102}$ & (1.191 $\pm$ 0.035 $^{+ 0.105}_{-0.104}$)  & (1.730 $\pm$ 0.143 $^{+ 0.266}_{-0.163}$) \\ 
0.095 - 0.10 & (0.759 $\pm$ 0.036 $^{+ 0.063}_{-0.096}$)		   & (1.111 $\pm$ 0.030 $^{+ 0.094}_{-0.098}$ & (1.332 $\pm$ 0.034 $^{+ 0.116}_{-0.100}$)  & (1.722 $\pm$ 0.156 $^{+ 0.263}_{-0.160}$) \\ 
0.10 - 0.11  & (0.818 $\pm$ 0.030 $^{+ 0.067}_{-0.093}$)		 & (1.151 $\pm$ 0.027 $^{+ 0.097}_{-0.094}$)  & (1.401 $\pm$ 0.031 $^{+ 0.121}_{-0.097}$)  & (2.016 $\pm$ 0.134 $^{+ 0.307}_{-0.158}$)   \\ 
0.11 - 0.12  & (0.893 $\pm$ 0.034 $^{+ 0.072}_{-0.089}$)		 & (1.230 $\pm$ 0.027 $^{+ 0.102}_{-0.091}$)  & (1.477 $\pm$ 0.031 $^{+ 0.126}_{-0.094}$)  & (2.210 $\pm$ 0.151 $^{+ 0.336}_{-0.157}$)   \\ 
0.12 - 0.13  & (0.928 $\pm$ 0.030 $^{+ 0.074}_{-0.086}$)		 & (1.334 $\pm$ 0.026 $^{+ 0.110}_{-0.088}$)  & (1.543 $\pm$ 0.029 $^{+ 0.131}_{-0.091}$)  & (2.162 $\pm$ 0.129 $^{+ 0.327}_{-0.155}$)   \\ 
0.13 - 0.15  & (0.960 $\pm$ 0.028 $^{+ 0.076}_{-0.084}$)		 & (1.411 $\pm$ 0.025 $^{+ 0.115}_{-0.086}$)  & (1.607 $\pm$ 0.027 $^{+ 0.135}_{-0.089}$)  & (2.256 $\pm$ 0.123 $^{+ 0.341}_{-0.154}$)   \\ 
0.15 - 0.16  & (1.030 $\pm$ 0.028 $^{+ 0.081}_{-0.083}$)		 & (1.453 $\pm$ 0.023 $^{+ 0.117}_{-0.085}$)  & (1.675 $\pm$ 0.026 $^{+ 0.139}_{-0.087}$)  & (2.447 $\pm$ 0.122 $^{+ 0.368}_{-0.153}$)   \\ 
0.16 - 0.17  & (1.085 $\pm$ 0.027 $^{+ 0.084}_{-0.081}$)		 & (1.477 $\pm$ 0.022 $^{+ 0.118}_{-0.083}$)  & (1.727 $\pm$ 0.025 $^{+ 0.142}_{-0.085}$)  & (2.359 $\pm$ 0.112 $^{+ 0.354}_{-0.152}$)   \\ 
0.17 - 0.19  & (1.123 $\pm$ 0.025 $^{+ 0.086}_{-0.079}$)		 & (1.545 $\pm$ 0.021 $^{+ 0.122}_{-0.081}$)  & (1.805 $\pm$ 0.024 $^{+ 0.147}_{-0.084}$)  & (2.277 $\pm$ 0.102 $^{+ 0.341}_{-0.151}$)   \\ 
0.19 - 0.21  & (1.164 $\pm$ 0.024 $^{+ 0.088}_{-0.078}$)		 & (1.613 $\pm$ 0.020 $^{+ 0.126}_{-0.080}$)  & (1.865 $\pm$ 0.023 $^{+ 0.150}_{-0.082}$)  & (2.423 $\pm$ 0.097 $^{+ 0.362}_{-0.150}$)   \\ 
0.21 - 0.22  & (1.194 $\pm$ 0.022 $^{+ 0.090}_{-0.076}$)		 & (1.634 $\pm$ 0.019 $^{+ 0.126}_{-0.078}$)  & (1.899 $\pm$ 0.021 $^{+ 0.152}_{-0.081}$)  & (2.592 $\pm$ 0.093 $^{+ 0.386}_{-0.149}$)   \\ 
0.22 - 0.24  & (1.222 $\pm$ 0.022 $^{+ 0.091}_{-0.075}$)		 & (1.673 $\pm$ 0.018 $^{+ 0.128}_{-0.077}$)  & (1.968 $\pm$ 0.021 $^{+ 0.156}_{-0.080}$)  & (2.647 $\pm$ 0.091 $^{+ 0.393}_{-0.149}$)   \\ 
0.24 - 0.26  & (1.278 $\pm$ 0.021 $^{+ 0.094}_{-0.074}$)		 & (1.732 $\pm$ 0.018 $^{+ 0.131}_{-0.076}$)  & (2.031 $\pm$ 0.020 $^{+ 0.159}_{-0.079}$)  & (2.682 $\pm$ 0.088 $^{+ 0.397}_{-0.148}$)   \\ 
0.26 - 0.29  & (1.359 $\pm$ 0.020 $^{+ 0.098}_{-0.073}$)		 & (1.775 $\pm$ 0.017 $^{+ 0.133}_{-0.075}$)  & (2.047 $\pm$ 0.019 $^{+ 0.159}_{-0.078}$)  & (2.684 $\pm$ 0.082 $^{+ 0.396}_{-0.148}$)   \\ 
0.29 - 0.31  & (1.371 $\pm$ 0.020 $^{+ 0.098}_{-0.072}$)		 & (1.788 $\pm$ 0.016 $^{+ 0.132}_{-0.074}$)  & (2.025 $\pm$ 0.018 $^{+ 0.155}_{-0.077}$)  & (2.603 $\pm$ 0.077 $^{+ 0.383}_{-0.147}$)   \\ 
0.31 - 0.34  & (1.380 $\pm$ 0.019 $^{+ 0.098}_{-0.071}$)		 & (1.818 $\pm$ 0.016 $^{+ 0.133}_{-0.073}$)  & (2.040 $\pm$ 0.017 $^{+ 0.155}_{-0.076}$)  & (2.656 $\pm$ 0.075 $^{+ 0.390}_{-0.147}$)   \\ 
0.34 - 0.36  & (1.381 $\pm$ 0.018 $^{+ 0.097}_{-0.070}$)		 & (1.822 $\pm$ 0.015 $^{+ 0.132}_{-0.072}$)  & (2.036 $\pm$ 0.017 $^{+ 0.153}_{-0.075}$)  & (2.633 $\pm$ 0.072 $^{+ 0.385}_{-0.146}$)   \\ 
0.36 - 0.39  & (1.398 $\pm$ 0.017 $^{+ 0.097}_{-0.069}$)		 & (1.816 $\pm$ 0.014 $^{+ 0.130}_{-0.072}$)  & (2.019 $\pm$ 0.016 $^{+ 0.150}_{-0.074}$)  & (2.504 $\pm$ 0.067 $^{+ 0.366}_{-0.146}$)   \\ 
0.39 - 0.43  & (1.410 $\pm$ 0.016 $^{+ 0.096}_{-0.068}$)		 & (1.778 $\pm$ 0.013 $^{+ 0.126}_{-0.071}$)  & (1.983 $\pm$ 0.015 $^{+ 0.146}_{-0.074}$)  & (2.415 $\pm$ 0.061 $^{+ 0.351}_{-0.146}$)   \\ 
0.43 - 0.46  & (1.396 $\pm$ 0.016 $^{+ 0.094}_{-0.068}$)		 & (1.754 $\pm$ 0.013 $^{+ 0.123}_{-0.070}$)  & (1.959 $\pm$ 0.014 $^{+ 0.143}_{-0.073}$)  & (2.495 $\pm$ 0.060 $^{+ 0.362}_{-0.145}$)   \\ 
0.46 - 0.50  & (1.391 $\pm$ 0.015 $^{+ 0.093}_{-0.067}$)		 & (1.715 $\pm$ 0.012 $^{+ 0.119}_{-0.069}$)  & (1.941 $\pm$ 0.014 $^{+ 0.140}_{-0.072}$)  & (2.383 $\pm$ 0.056 $^{+ 0.345}_{-0.145}$)   \\ 
0.50 - 0.54  & (1.349 $\pm$ 0.014 $^{+ 0.089}_{-0.066}$)		 & (1.677 $\pm$ 0.012 $^{+ 0.115}_{-0.068}$)  & (1.844 $\pm$ 0.013 $^{+ 0.132}_{-0.071}$)  & (2.200 $\pm$ 0.052 $^{+ 0.318}_{-0.144}$)   \\ 
0.54 - 0.58  & (1.328 $\pm$ 0.014 $^{+ 0.086}_{-0.065}$)		 & (1.643 $\pm$ 0.011 $^{+ 0.111}_{-0.068}$)  & (1.789 $\pm$ 0.012 $^{+ 0.126}_{-0.071}$)  & (2.213 $\pm$ 0.050 $^{+ 0.319}_{-0.144}$)   \\ 
0.58 - 0.62  & (1.300 $\pm$ 0.009 $^{+ 0.084}_{-0.064}$)		 & (1.577 $\pm$ 0.010 $^{+ 0.105}_{-0.067}$)  & (1.731 $\pm$ 0.013 $^{+ 0.121}_{-0.070}$)  & (2.107 $\pm$ 0.032 $^{+ 0.303}_{-0.144}$)   \\ 
0.62 - 0.67  & (1.254 $\pm$ 0.008 $^{+ 0.080}_{-0.064}$)		 & (1.527 $\pm$ 0.009 $^{+ 0.101}_{-0.066}$)  & (1.641 $\pm$ 0.012 $^{+ 0.114}_{-0.069}$)  & (1.966 $\pm$ 0.030 $^{+ 0.282}_{-0.143}$)   \\ 
0.67 - 0.72  & (1.220 $\pm$ 0.008 $^{+ 0.077}_{-0.063}$)		 & (1.468 $\pm$ 0.009 $^{+ 0.096}_{-0.065}$)  & (1.587 $\pm$ 0.011 $^{+ 0.109}_{-0.068}$)  & (1.870 $\pm$ 0.028 $^{+ 0.268}_{-0.143}$)   \\ 
0.72 - 0.77  & (1.173 $\pm$ 0.007 $^{+ 0.073}_{-0.062}$)		 & (1.403 $\pm$ 0.008 $^{+ 0.091}_{-0.065}$)  & (1.510 $\pm$ 0.011 $^{+ 0.102}_{-0.068}$)  & (1.769 $\pm$ 0.026 $^{+ 0.252}_{-0.143}$)   \\ 
0.77 - 0.83  & (1.125 $\pm$ 0.007 $^{+ 0.069}_{-0.061}$)		 & (1.341 $\pm$ 0.007 $^{+ 0.086}_{-0.064}$)  & (1.431 $\pm$ 0.010 $^{+ 0.096}_{-0.067}$)  & (1.710 $\pm$ 0.025 $^{+ 0.243}_{-0.142}$)   \\ 
0.83 - 0.89  & (1.064 $\pm$ 0.006 $^{+ 0.064}_{-0.060}$)		 & (1.272 $\pm$ 0.007 $^{+ 0.080}_{-0.063}$)  & (1.357 $\pm$ 0.009 $^{+ 0.090}_{-0.066}$)  & (1.550 $\pm$ 0.022 $^{+ 0.220}_{-0.142}$)   \\ 
0.89 - 0.96  & (1.003 $\pm$ 0.006 $^{+ 0.060}_{-0.060}$)		 & (1.197 $\pm$ 0.007 $^{+ 0.075}_{-0.062}$)  & (1.273 $\pm$ 0.009 $^{+ 0.084}_{-0.066}$)  & (1.442 $\pm$ 0.021 $^{+ 0.204}_{-0.142}$)   \\ 
0.96 - 1.02  & (0.964 $\pm$ 0.005 $^{+ 0.057}_{-0.059}$)		 & (1.135 $\pm$ 0.006 $^{+ 0.070}_{-0.062}$)  & (1.192 $\pm$ 0.008 $^{+ 0.077}_{-0.065}$)  & (1.382 $\pm$ 0.020 $^{+ 0.195}_{-0.141}$)   \\ 
1.02 - 1.09  & (0.919 $\pm$ 0.005 $^{+ 0.053}_{-0.058}$)		 & (1.066 $\pm$ 0.006 $^{+ 0.065}_{-0.061}$)  & (1.127 $\pm$ 0.007 $^{+ 0.072}_{-0.064}$)  & (1.294 $\pm$ 0.018 $^{+ 0.183}_{-0.141}$)   \\ 
1.09 - 1.17  & (0.860 $\pm$ 0.005 $^{+ 0.049}_{-0.057}$)		 & (0.993 $\pm$ 0.005 $^{+ 0.060}_{-0.060}$)  & (1.041 $\pm$ 0.007 $^{+ 0.066}_{-0.064}$)  & (1.178 $\pm$ 0.017 $^{+ 0.166}_{-0.141}$)   \\ 
1.17 - 1.25  & (0.813 $\pm$ 0.004 $\pm$ 0.046)  			 & (0.934 $\pm$ 0.005 $\pm$ 0.056)	      & (0.983 $\pm$ 0.006 $\pm$ 0.062) 	    & (1.105 $\pm$ 0.016 $\pm$ 0.155) \\ 
1.25 - 1.34  & (0.753 $\pm$ 0.004 $\pm$ 0.042)  			 & (0.860 $\pm$ 0.005 $\pm$ 0.051)	      & (0.903 $\pm$ 0.006 $\pm$ 0.056) 	    & (1.030 $\pm$ 0.014 $\pm$ 0.144) \\ 
1.34 - 1.42  & (0.708 $\pm$ 0.004 $\pm$ 0.039)  			 & (0.813 $\pm$ 0.004 $\pm$ 0.047)	      & (0.845 $\pm$ 0.006 $\pm$ 0.052) 	    & (0.931 $\pm$ 0.013 $\pm$ 0.130) \\ 
1.42 - 1.52  & (0.650 $\pm$ 0.004 $\pm$ 0.035)  			 & (0.748 $\pm$ 0.004 $\pm$ 0.043)	      & (0.782 $\pm$ 0.005 $\pm$ 0.048) 	    & (0.862 $\pm$ 0.012 $\pm$ 0.120) \\ 
1.52 - 1.62  & (0.607 $\pm$ 0.003 $\pm$ 0.033)  			 & (0.693 $\pm$ 0.004 $\pm$ 0.039)	      & (0.719 $\pm$ 0.005 $\pm$ 0.044) 	    & (0.789 $\pm$ 0.011 $\pm$ 0.110) \\ 
1.62 - 1.72  & (0.562 $\pm$ 0.003 $\pm$ 0.030)  			 & (0.640 $\pm$ 0.003 $\pm$ 0.036)	      & (0.670 $\pm$ 0.004 $\pm$ 0.040) 	    & (0.720 $\pm$ 0.010 $\pm$ 0.100) \\ 
1.72 - 1.83  & (0.520 $\pm$ 0.003 $\pm$ 0.027)  			 & (0.590 $\pm$ 0.003 $\pm$ 0.033)	      & (0.612 $\pm$ 0.004 $\pm$ 0.036) 	    & (0.675 $\pm$ 0.010 $\pm$ 0.094) \\ 
1.83 - 1.95  & (0.481 $\pm$ 0.003 $\pm$ 0.025)  			 & (0.540 $\pm$ 0.003 $\pm$ 0.030)	      & (0.562 $\pm$ 0.004 $\pm$ 0.033) 	    & (0.605 $\pm$ 0.009 $\pm$ 0.084) \\ 
1.95 - 2.07  & (0.441 $\pm$ 0.002 $\pm$ 0.023)  			 & (0.501 $\pm$ 0.003 $\pm$ 0.027)	      & (0.512 $\pm$ 0.004 $\pm$ 0.030) 	    & (0.547 $\pm$ 0.008 $\pm$ 0.076) \\ 
2.07 - 2.20  & (0.412 $\pm$ 0.002 $\pm$ 0.021)  			 & (0.450 $\pm$ 0.002 $\pm$ 0.024)	      & (0.461 $\pm$ 0.003 $\pm$ 0.027) 	    & (0.499 $\pm$ 0.008 $\pm$ 0.069) \\ 
2.20 - 2.33  & (0.374 $\pm$ 0.002 $\pm$ 0.019)  			 & (0.409 $\pm$ 0.002 $\pm$ 0.022)	      & (0.428 $\pm$ 0.003 $\pm$ 0.024) 	    & (0.450 $\pm$ 0.007 $\pm$ 0.062) \\ 
2.33 - 2.48  & (0.347 $\pm$ 0.002 $\pm$ 0.017)  			 & (0.377 $\pm$ 0.002 $\pm$ 0.020)	      & (0.388 $\pm$ 0.003 $\pm$ 0.022) 	    & (0.411 $\pm$ 0.006 $\pm$ 0.057) \\ 
2.48 - 2.62  & (0.315 $\pm$ 0.002 $\pm$ 0.015)  			 & (0.348 $\pm$ 0.002 $\pm$ 0.018)	      & (0.356 $\pm$ 0.003 $\pm$ 0.020) 	    & (0.390 $\pm$ 0.006 $\pm$ 0.054) \\ 
2.62 - 2.78  & (0.286 $\pm$ 0.002 $\pm$ 0.014)  			 & (0.318 $\pm$ 0.002 $\pm$ 0.016)	      & (0.322 $\pm$ 0.002 $\pm$ 0.018) 	    & (0.334 $\pm$ 0.005 $\pm$ 0.046) \\ 
2.78 - 2.94  & (0.262 $\pm$ 0.002 $\pm$ 0.012)  			 & (0.288 $\pm$ 0.002 $\pm$ 0.015)	      & (0.291 $\pm$ 0.002 $\pm$ 0.016) 	    & (0.303 $\pm$ 0.005 $\pm$ 0.042) \\ 
2.94 - 3.12  & (0.239 $\pm$ 0.001 $\pm$ 0.011)  			 & (0.258 $\pm$ 0.002 $\pm$ 0.013)	      & (0.262 $\pm$ 0.002 $\pm$ 0.014) 	    & (0.276 $\pm$ 0.005 $\pm$ 0.038) \\ 
3.12 - 3.30  & (0.215 $\pm$ 0.001 $\pm$ 0.010)  			 & (0.234 $\pm$ 0.001 $\pm$ 0.012)	      & (0.239 $\pm$ 0.002 $\pm$ 0.013) 	    & (0.251 $\pm$ 0.004 $\pm$ 0.034) \\ 
3.30 - 3.49  & (0.195 $\pm$ 0.001 $\pm$ 0.009)  			 & (0.210 $\pm$ 0.001 $\pm$ 0.010)	      & (0.215 $\pm$ 0.002 $\pm$ 0.012) 	    & (0.223 $\pm$ 0.004 $\pm$ 0.031) \\ 
3.49 - 3.69  & (0.175 $\pm$ 0.001 $\pm$ 0.008)  			 & (0.189 $\pm$ 0.001 $\pm$ 0.009)	      & (0.194 $\pm$ 0.002 $\pm$ 0.010) 	    & (0.204 $\pm$ 0.004 $\pm$ 0.028) \\ 

3.69 - 4.12  & (151.4 $\pm$ 0.7 $\pm$ 6.9) $\times$ 10$^{-3}$       & (163.3 $\pm$ 0.7565 $\pm$ 8.050) $\times$ 10$^{-3}$   & (164.7 $\pm$ 0.9853 $\pm$ 8.790)  $\times$ 10$^{-3}$ & (0.172 $\pm$ 0.002 $\pm$ 0.023) 			  \\ 
4.12 - 4.59  & (124.4 $\pm$ 0.6 $\pm$ 5.6) $\times$ 10$^{-3}$       & (132.0 $\pm$ 0.6391 $\pm$ 6.446) $\times$ 10$^{-3}$   & (134.3 $\pm$ 0.8364 $\pm$ 7.113)  $\times$ 10$^{-3}$ & (0.141 $\pm$ 0.002 $\pm$ 0.019) 			  \\ 
4.59 - 5.11  & (99.15 $\pm$ 0.5 $\pm$ 4.4) $\times$ 10$^{-3}$       & (106.8 $\pm$ 0.5402 $\pm$ 5.168) $\times$ 10$^{-3}$   & (109.1 $\pm$ 0.7083 $\pm$ 5.731)  $\times$ 10$^{-3}$ & (0.113 $\pm$ 0.002 $\pm$ 0.015) 			  \\ 
5.11 - 5.68  & (81.38 $\pm$ 0.4 $\pm$ 3.6) $\times$ 10$^{-3}$      & (85.34 $\pm$ 0.4537 $\pm$ 4.093)  $\times$ 10$^{-3}$  & (86.97 $\pm$ 0.5944 $\pm$ 4.536)  $\times$ 10$^{-3}$  & (0.091 $\pm$ 0.001 $\pm$ 0.012) 			  \\ 
5.68 - 6.30  & (65.27 $\pm$ 0.4 $\pm$ 2.9) $\times$ 10$^{-3}$      & (68.61 $\pm$ 0.3820 $\pm$ 3.262)  $\times$ 10$^{-3}$  & (69.55 $\pm$ 0.4994 $\pm$ 3.601)  $\times$ 10$^{-3}$  & (0.069 $\pm$ 0.001 $\pm$ 0.009) 			  \\ 
6.30 - 6.99  & (52.12 $\pm$ 0.3 $\pm$ 2.3) $\times$ 10$^{-3}$      & (54.86 $\pm$ 0.3211 $\pm$ 2.587)  $\times$ 10$^{-3}$  & (55.60 $\pm$ 0.4197 $\pm$ 2.859)  $\times$ 10$^{-3}$  &	(56.69 $\pm$ 0.9533 $\pm$ 7.693)   $\times$ 10$^{-3}$\\ 
6.99 - 7.74  & (41.66 $\pm$ 0.3 $\pm$ 1.8) $\times$ 10$^{-3}$      & (43.65 $\pm$ 0.2691 $\pm$ 2.042)  $\times$ 10$^{-3}$  & (44.15 $\pm$ 0.3514 $\pm$ 2.255)  $\times$ 10$^{-3}$  &	(44.91 $\pm$ 0.7967 $\pm$ 6.088)   $\times$ 10$^{-3}$\\ 
7.74 - 8.57  & (33.18 $\pm$ 0.2 $\pm$ 1.4) $\times$ 10$^{-3}$      & (34.31 $\pm$ 0.2241 $\pm$ 1.593)  $\times$ 10$^{-3}$  & (34.65 $\pm$ 0.2923 $\pm$ 1.759)  $\times$ 10$^{-3}$  &	(34.70 $\pm$ 0.6572 $\pm$ 4.700)   $\times$ 10$^{-3}$\\ 
8.57 - 9.48  & (26.26 $\pm$ 0.2 $\pm$ 1.1) $\times$ 10$^{-3}$      & (26.80 $\pm$ 0.1856 $\pm$ 1.235)  $\times$ 10$^{-3}$  & (27.12 $\pm$ 0.2424 $\pm$ 1.368)  $\times$ 10$^{-3}$  &	(27.44 $\pm$ 0.5473 $\pm$ 3.713)   $\times$ 10$^{-3}$\\ 

9.48 - 10.48  &  (208.8 $\pm$ 1.5 $\pm$ 8.8) $\times$ 10$^{-4}$    &	 (214.3 $\pm$ 1.556 $\pm$ 9.872)  $\times$ 10$^{-4}$  & (21.32 $\pm$ 0.2014 $\pm$ 1.075)  $\times$ 10$^{-3}$	&   (21.77 $\pm$ 0.4564 $\pm$ 2.945)   $\times$ 10$^{-3}$\\ 
10.48 - 11.57  & (164.0 $\pm$ 1.2 $\pm$ 6.9) $\times$ 10$^{-4}$    & (167.1 $\pm$ 1.286 $\pm$ 7.697)  $\times$ 10$^{-4}$       & (167.7 $\pm$ 1.672 $\pm$ 8.456)  $\times$ 10$^{-4}$	& (16.86 $\pm$ 0.3755 $\pm$ 2.281)   $\times$ 10$^{-3}$\\   
11.57 - 12.77  & (128.9 $\pm$ 1.0 $\pm$ 5.5) $\times$ 10$^{-4}$    & (130.2 $\pm$ 1.061 $\pm$ 5.998)  $\times$ 10$^{-4}$      & (134.6 $\pm$ 1.399 $\pm$ 6.785)  $\times$ 10$^{-4}$	& (13.03 $\pm$ 0.3083 $\pm$ 1.763)   $\times$ 10$^{-3}$\\   
12.77 - 14.09  & (100.1 $\pm$ 0.8 $\pm$ 4.2) $\times$ 10$^{-4}$    & (103.4 $\pm$ 0.8845 $\pm$ 4.762)  $\times$ 10$^{-4}$    & (102.8 $\pm$ 1.143 $\pm$ 5.182)  $\times$ 10$^{-4}$	& (10.22 $\pm$ 0.2554 $\pm$ 1.384)   $\times$ 10$^{-3}$\\   
14.09 - 15.54  & (78.59 $\pm$ 0.7 $\pm$ 3.3) $\times$ 10$^{-4}$    & (79.72 $\pm$ 0.7250 $\pm$ 3.672)   $\times$ 10$^{-4}$   & (78.55 $\pm$ 0.9319 $\pm$ 3.960)  $\times$ 10$^{-4}$	& (8.339 $\pm$ 0.2150 $\pm$ 1.129)   $\times$ 10$^{-3}$ \\  
15.54 - 17.12  & (62.71 $\pm$ 0.6 $\pm$ 2.7) $\times$ 10$^{-4}$    & (64.08 $\pm$ 0.6064 $\pm$ 2.952)   $\times$ 10$^{-4}$   & (62.57 $\pm$ 0.7736 $\pm$ 3.154)  $\times$ 10$^{-4}$	& (62.71 $\pm$ 1.735 $\pm$ 8.486)  $\times$ 10$^{-4}$ \\    
17.12 - 18.86  & (47.92 $\pm$ 0.5 $\pm$ 2.0) $\times$ 10$^{-4}$    & (49.08 $\pm$ 0.4937 $\pm$ 2.261)   $\times$ 10$^{-4}$   & (49.15 $\pm$ 0.6357 $\pm$ 2.478)  $\times$ 10$^{-4}$	& (47.39 $\pm$ 1.400 $\pm$ 6.413) $\times$ 10$^{-4}$ \\     
18.86 - 20.76  & (37.16 $\pm$ 0.4 $\pm$ 1.6) $\times$ 10$^{-4}$    & (37.96 $\pm$ 0.4066 $\pm$ 1.748)   $\times$ 10$^{-4}$   & (38.06 $\pm$ 0.5230 $\pm$ 1.919)  $\times$ 10$^{-4}$	& (38.42 $\pm$ 1.181 $\pm$ 5.200) $\times$ 10$^{-4}$ \\     
20.76 - 22.85  & (28.92 $\pm$ 0.3 $\pm$ 1.2) $\times$ 10$^{-4}$    & (29.32 $\pm$ 0.3398 $\pm$ 1.351)   $\times$ 10$^{-4}$   & (28.47 $\pm$ 0.4282 $\pm$ 1.435)  $\times$ 10$^{-4}$	& (28.76 $\pm$ 0.9688 $\pm$ 3.892) $\times$ 10$^{-4}$ \\     

22.85 - 25.15  & (225.8 $\pm$ 2.6 $\pm$ 9.6) $\times$ 10$^{-5}$     & (23.07 $\pm$ 0.2879 $\pm$ 1.063)   $\times$ 10$^{-4}$ & (22.61 $\pm$ 0.3642 $\pm$ 1.140)  $\times$ 10$^{-4}$   & (21.59 $\pm$ 0.8009 $\pm$ 2.922) $\times$ 10$^{-4}$ \\ 
25.15 - 27.66  & (171.3 $\pm$ 2.2 $\pm$ 7.2) $\times$ 10$^{-5}$     & (173.8 $\pm$ 2.389 $\pm$ 8.006)   $\times$ 10$^{-5}$   & (173.5 $\pm$ 3.049 $\pm$ 8.745)  $\times$ 10$^{-5}$    & (18.35 $\pm$ 0.7055 $\pm$ 2.484) $\times$ 10$^{-4}$\\ 
27.66 - 30.42  & (138.0 $\pm$ 1.9 $\pm$ 5.8) $\times$ 10$^{-5}$     & (137.7 $\pm$ 2.031 $\pm$ 6.344)  $\times$ 10$^{-5}$    & (138.8 $\pm$ 2.603 $\pm$ 6.997)  $\times$ 10$^{-5}$    & (14.03 $\pm$ 0.5883 $\pm$ 1.898) $\times$ 10$^{-4}$\\ 
30.42 - 33.44  & (103.8 $\pm$ 1.5 $\pm$ 4.4) $\times$ 10$^{-5}$     & (105.6 $\pm$ 1.699 $\pm$ 4.862)  $\times$ 10$^{-5}$    & (101.5 $\pm$ 2.127 $\pm$ 5.118)  $\times$ 10$^{-5}$    & (10.95 $\pm$ 0.4965 $\pm$ 1.482) $\times$ 10$^{-4}$\\ 
33.44 - 36.75  & (81.42 $\pm$ 1.3 $\pm$ 3.4) $\times$ 10$^{-5}$     & (80.24 $\pm$ 1.416 $\pm$ 3.696)  $\times$ 10$^{-5}$    & (81.51 $\pm$ 1.820 $\pm$ 4.109)  $\times$ 10$^{-5}$    & (76.20 $\pm$ 3.955 $\pm$ 10.31) $\times$ 10$^{-5}$\\ 
36.75 - 40.39  & (63.41 $\pm$ 1.1 $\pm$ 2.7) $\times$ 10$^{-5}$     & (62.97 $\pm$ 1.199 $\pm$ 2.900)  $\times$ 10$^{-5}$    & (62.22 $\pm$ 1.520 $\pm$ 3.137)  $\times$ 10$^{-5}$    & (61.51 $\pm$ 3.396 $\pm$ 8.323)  $\times$ 10$^{-5}$\\ 
40.39 - 44.37  & (47.29 $\pm$ 0.9 $\pm$ 2.0) $\times$ 10$^{-5}$    & (49.30 $\pm$ 1.014 $\pm$ 2.271)  $\times$ 10$^{-5}$    & (49.86 $\pm$ 1.300 $\pm$ 2.514)  $\times$ 10$^{-5}$     & (47.38 $\pm$ 2.847 $\pm$ 6.412)  $\times$ 10$^{-5}$\\ 
44.37 - 48.74  & (35.22 $\pm$ 0.8 $\pm$ 1.5) $\times$ 10$^{-5}$    & (33.67 $\pm$ 0.801 $\pm$ 1.551)  $\times$ 10$^{-5}$    & (35.50 $\pm$ 1.049 $\pm$ 1.790)  $\times$ 10$^{-5}$     & (36.99 $\pm$ 2.404 $\pm$ 5.006)  $\times$ 10$^{-5}$\\ 

\enddata

\end{deluxetable}

\end{document}